\documentclass{article}

\usepackage[preprint]{neurips_2024}

\usepackage[utf8]{inputenc} 
\usepackage[T1]{fontenc}    
\usepackage{hyperref}       
\usepackage{url}            
\usepackage{booktabs}       
\usepackage{amsfonts}       
\usepackage{nicefrac}       
\usepackage{microtype}      
\usepackage{xcolor}         
\usepackage{graphicx}
\usepackage{amsmath} 
\usepackage{tikz}
\usetikzlibrary{arrows.meta}
\usepackage{multirow}
\usepackage{makecell}
\usepackage{algorithm}
\usepackage{algorithmic}

\usepackage{amsmath}
\usepackage{amssymb}
\usepackage{mathtools}
\usepackage{amsthm}
\usepackage{mathrsfs}

\title{Hierarchical Reinforcement Learning for Temporal Abstraction of Listwise Recommendation}

\author{%
  Luo Ji \thanks{Corresponding author: \texttt{jiluo.lj@alibaba-inc.com}}, Gao Liu, Mingyang Yin, Hongxia Yang, Jingren Zhou \\ 
  DAMO Academy, Alibaba Group 
}
\begin{document}

\maketitle

\begin{abstract}
Modern listwise recommendation systems need to consider both long-term user perceptions and short-term interest shifts. Reinforcement learning can be applied on recommendation to study such a problem but is also subject to large search space, sparse user feedback and long interactive latency. Motivated by recent progress in hierarchical reinforcement learning, we propose a novel framework called mccHRL to provide different levels of temporal abstraction on listwise recommendation. Within the hierarchical framework, the high-level agent studies the evolution of user perception, while the low-level agent produces the item selection policy by modeling the process as a sequential decision-making problem. We argue that such framework has a well-defined decomposition of the outra-session context and the intra-session context, which are encoded by the high-level and low-level agents, respectively. To verify this argument, we implement both a simulator-based environment and an industrial dataset-based experiment. Results observe significant performance improvement by our method, compared with several well-known baselines. Data and codes have been made public.
\end{abstract}

\section{Introduction}

In recent years, hybrid Recommender systems (RS) have become increasingly popular to overcome information redundancy and have been widely utilized in a variety of domains (\emph{e.g.} products, articles, advertisements, music, and movies) ~\citep{Anidorif2015Recommender}. Hybrid RS selects a list of contents (usually called `items') from overwhelmed candidates for exhibition to meet user preferences, with expected clicks, dwell times or purchases increased. Although with substantial achievement, more advanced recommendation techniques are always necessary because of the complexity of business scenarios, including listwise, cold-start, heterogeneous contents, spatiotemporal-aware (STA) recommendations, etc. Among these, the listwise recommendation is always an important issue, not only due to the wide application of Top-K recommendation ~\citep{Chen2019TopKOC} or feed recommendation ~\citep{Wu2021FeedRec}, but also the computational complexity with consideration of mutual-item influence, and position bias.



\begin{figure}[t]
  \centering
  \includegraphics[width=0.9\columnwidth]{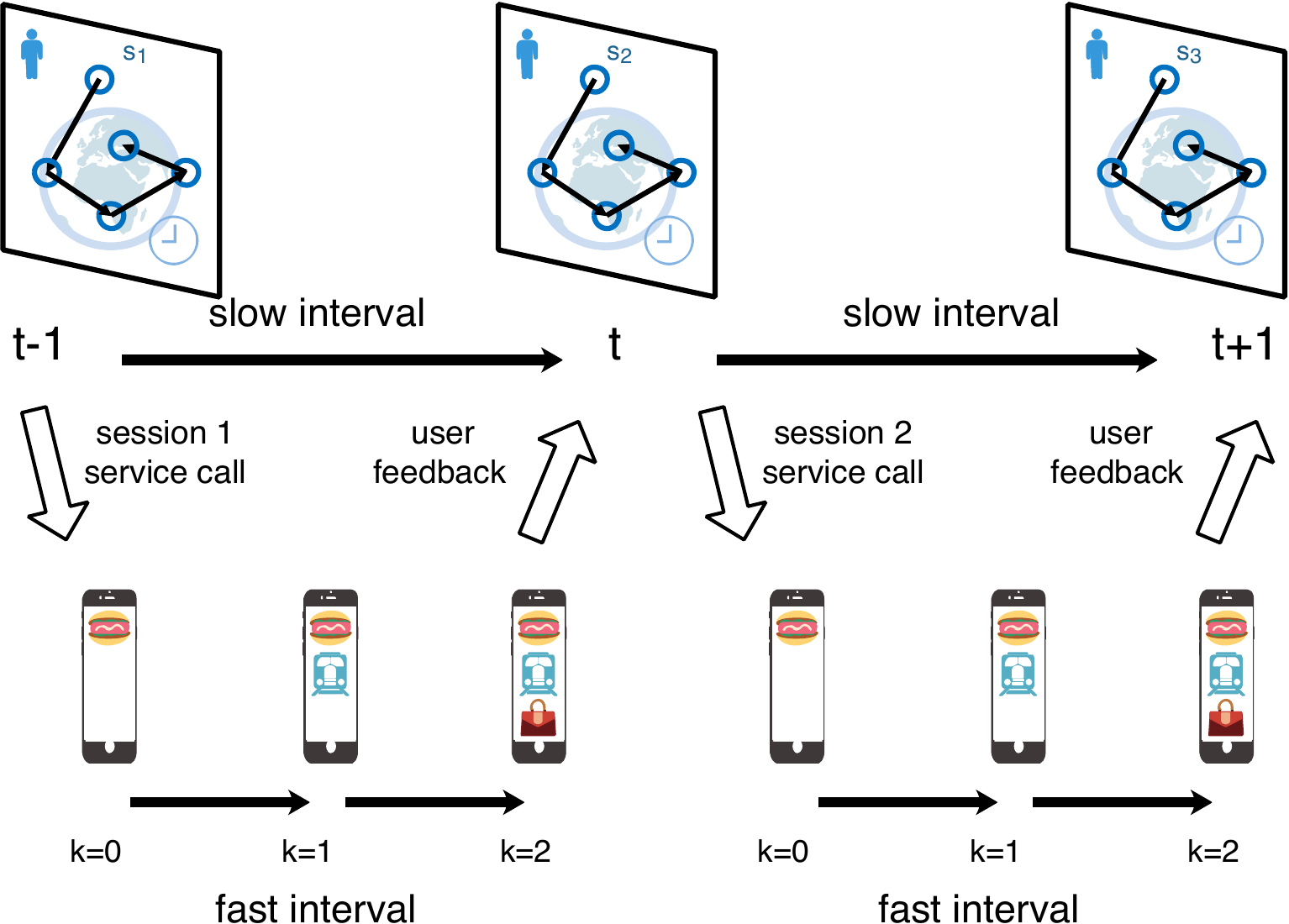} 
  \caption{Two levels of temporal abstractions in typical listwise recommendations. The user requests a session service at time $t$, then a ranking policy is executed $k$ steps to provide a top-k recommendation.}
  \label{fig:environment}
\end{figure}

Reinforcement Learning (RL) is a natural, unified framework to learn interactively with the environment and maximize the global and long-term rewards, which highlights some promising solutions of listwise ranking. According to the formulation manner of the Markov Decision Process (MDP), we conclude that most such works can be classified into two main categories: 
\begin{enumerate}
    \item Define the user perception/preference as state and the entire recommendation list as action, and model the user interest shift as state transition ~\citep{Zheng2018DRN,Chen2018StabilizingRL,Chen2019TopKOC,Ie2019SLATEQ}.
    \item Formulate the listwise ranking problem as a sequential decision-making process, with the state represented by the current list ranked item, while the action is to select the next optimal item to the list \citep{Gong2019ExactKRV,zhao2019LIRD,Zhao2018DeepPage}.
\end{enumerate}
Nevertheless, both methodologies have their obstacles, which prevent the wide application of RL on industrial RS. For the first category, a thorough representation of user states can be expected especially for STA or POI-based recommendations, but the modeling of state transition is questionable since its modeling time interval is coarse-grained. Users naturally interact with RS based on the item-wise experience instead of the session, and they might leave the app for a while before the next session experience, both of which make the Markov assumption questionable. The second category, on the other hand, can consider the mutual influence of items, the intra-session user interest shift, and may even interact with users within the session. However, it is subject to the curse of dimensionality when serving all users simultaneously at a centralized server. It also has the sparse reward issue for an RL framework since the user response can not be observed by the agent until the end of the session. Figure \ref{fig:environment} indicates these two temporal patterns of listwise ranking.

Here we aim to overcome the above obstacles through better temporal abstractions, by designing a fast-slow learning paradigm \citep{Madan2021metaRim} of hierarchical reinforcement learning (HRL). Although there are some previous HRL-based RS efforts  \citep{Takanobu2019heteroHRL, Zhao2020MaHRL, Xie2021HRL-Rec}, they might focus on heterogeneous or multi-goal recommendation problems. Instead, we design our HRL such that one level encodes the user perception and outer-session context, while another level encodes the spatiotemporal status, intra-session context, and mutual-item influences. Besides that, we further improve the sample efficiency by applying Edge Computing or Edge-based AI \citep{zhou2019edge} on our RS. With the rapid development of Edge Computing or Edge AI \citep{zhou2019edge}, it is possible to utilize user features and feedback on mobile \citep{Gong2020EdgeRec} or collaborative training with cloud \citep{yao2021dccl, chen2021mc2sf} in recommendation. We embrace this benefit to further improve the modeling depth of user states and enhance the Markov assumption. We argue that the low-level HRL could be deployed on mobile devices, therefore the on-device features can be involved, training is decoupled, and the communication frequency of cloud service is reduced during model inference. Although this intuition belongs to the paradigm of collaborative training with mobile and cloud \citep{yao2021dccl, chen2021mc2sf}, however, it is the first attempt that tries to train part of RL components on the edge side, to the best of our knowledge \citep{yao2022edgecloud}. 


In this paper, we propose a novel methodology, named \textit{mobile-cloud collaborative  Hierarchical Reinforcement Learning} (mccHRL)  for Listwise Recommendation, which employs HRL to solve a highly STA, listwise RS problem on a world-leading mobile application. The High-Level of mccHRL models the user perception and STA status, interacts with user responses at the end of session, and takes the suggested long-term user preference as action; while the Low-Level of mccHRL solves the listwise ranking problem, by providing an item selection policy based on user short-term interest and on-device features. The High-Level action is utilized as the goal of Low-Level policy to achieve, such that the Low-Level agent does not directly interact with user response. Embedding studied on the High-Level are transmitted to the Low-Level policy for knowledge transfer and generalization. Detailed implementation has been made public\footnote{https://anonymous.4open.science/r/eccHRL-F99B/}. To conclude, the major contributions of this paper include:



\begin{itemize}
\item We propose a novel HRL framework on listwise recommendation with a natural way of temporal abstractions. The High-Level agent studies the outra-session context (spatiotemporal effects) and directly interacts with users, while the Low-Level agent studies the intra-session context, intrinsically motivated by the high-level. 


\item We provide an explicit and detailed implementation of on-device features and cloud transmission latency in our framework. The HRL solves the mobile-cloud collaboration with its hierarchical architecture.


\item Before the online deployment, we design two offline pipelines, including simulator-based and dataset-based experiments. We also develop the offline version of training algorithm and its performance is verified by industrial-scale experiments.

\item We design and implement the realistic mobile-cloud recommendation environment in our experiments. On-device features, user one-device feedback, and mobile-cloud transmission delays are explicitly considered in an online studying manner, which is seldom practiced by previous edge-based recommendation studies.
\end{itemize}


The rest of the paper is organized as follows. The connection with previous works is first discussed in Section \ref{sec:related_work}. Preliminaries and the problem formulation are then introduced in Section \ref{sec:preliminaries}. Our methodology and corresponding algorithms are stated in Section \ref{sec:method}. Experiment results are summarized in Section \ref{sec:experiment}. Finally Section \ref{sec:conclusion} concludes this paper.


\section{Related Work}
\label{sec:related_work}

\subsection{RL-based Recommendation}
\label{seq:rl-rec}

RL has been widely used in recommendation tasks, while their definitions of MDP can be diversified into two categories. The first categories models the user preference as state and recommendation the entire item list as action, such as \citep{Zheng2018DRN,Chen2018StabilizingRL,Chen2019TopKOC,Ie2019SLATEQ}, in which the state transit depicts the user interest shift. On the other hand, the second category is to model the ranking steps of items as state, and the selection of the next favorite item is action. Such methods include MDPrank \citep{Xia2017MDPDIV,Wei2017MDPRank}, Exact-K \citep{Gong2019ExactKRV}, LIRD \citep{zhao2019LIRD} and DeepPage \citep{Zhao2018DeepPage}, which aims to model the mutual influence between items and listwise bias. The intuition of combining the advantages of two types of MDPs motivates our idea of mccHRL.

There are also RL works aiming to fix some special recommendation issues, such as location or POI-based information \citep{Zhou2019APOIR}, cold start \citep{Wang2020M3Rec}, heterogeneous items \citep{Takanobu2019heteroHRL, Xie2021HRL-Rec}, user long-term engagement \citep{2019UserEngage} or fairness of items \citep{Ge2021FCPO}. Our work shares similar interests to the POI-based recommendation with \citep{lian2020GeoSAN, Zhou2019APOIR, Luo2021STAN}. However, most of them work on the next-location recommendation given sparse spatial information, to learn a reasonable user-location relationship matrix, while we aim to provide an end-to-end solution of session recommendation with spatial information implicitly considered in our recommending policy.

\subsection{Hierarchical Reinforcement Learning}

Hierarchical Reinforcement Learning (HRL) has a hierarchical structure of RL layers in order to solve more complex tasks, reduce the searching space,  provide different levels of temporal abstractions, or deal with sparse rewards. HRL methods can be classified into two categories by the coordinated way between different levels. The first is the goal-conditional framework \citep{Kulkarni2016h-DQN, Vezhnevets2017FuN, Nachum2018HIRO, Levy2019HAC}, in which a high-level agent learns the goal to drive the low-level agent; the second is sub-task discovery framework, the high-level agent might provide some options (or skill) to reduce the search space of low-level agent \citep{Bacon2017Option-Critic, Florensa2017snn4hrl}. Our methodology belongs to the first category in which the high-level agent studies the user perception embedding as the low-level agent decision basis.

HRL methods also vary in hierarchical structures. Choices include multi-layer DQN \citep{Kulkarni2016h-DQN}, multi-layer policy \citep{Nachum2018HIRO}, or multi-layer actor-critic \citep{Levy2019HAC}. Our model is similar to hierarchical actor-critic (HAC) but reduces its complexity for practical edge deployment.


\subsection{HRL-based Recommendation}

There are also attempts to apply Hierarchical Reinforcement Learning (HRL) on recommendation or search \citep{Takanobu2019heteroHRL,Zhao2020MaHRL,Xie2021HRL-Rec}. \citep{Takanobu2019heteroHRL} uses the High-Level RL as the heterogeneous search source selector; similarly, High-Level RL in HRL-rec \citep{Xie2021HRL-Rec} is the content channel selector. MaHRL \citep{Zhao2020MaHRL} uses the High-Level RL to model the user perception state, to optimize multiple recommendation goals. All of these methods define the Low-Level RL as the item selector, which is similar to our framework. We have a similar definition of High-Level RL with MaHRL, however, our difference includes that (1) our target is to improve a session-based, STA recommendation CTR performance instead of multiple labels; (2) we provide a natural formulation of temporal abstraction to solve the sparse reward issue of listwise recommendation; (3) we deploy the Low-Level part of HRL on the edge side to further improve the methodology throughput. 


\subsection{Edge-based Recommendation}

Edge Computing (might also be named Edge Intelligence, Edge AI, or on-device AI), in contrast to Cloud computing, has been widely studied recently years \citep{zhou2019edge}. Efforts have been made on applications including IoT, 5G, and auto-driving. However, this field still is at its early stage with most efforts focusing on the lightweight and deployment of edge models \citep{yao2022edgecloud}. For example, EdgeRec \citep{Gong2020EdgeRec} works on the split-deployment which places the memory-consuming embedding module on the cloud while the lightweight recommender is inferenced on the device.  

There has been increasing attention on Edge-Cloud collaboration, either privacy-primary such as Federated Learning, or efficient-primary. We are interested in efficient primary methods, to improve the recommender's personalization. Such efforts including COLLA \citep{Lu2019COLLA}, DCCL \citep{yao2021dccl} and MC$^2$-SF \citep{chen2021mc2sf}. For example, COLLA designed the cloud model as a global aggregator distilling knowledge from many edge models; MC$^2$-SF proposes  a slow-fast learning mechanism on RS, with the bidirectional collaboration between Edge and Cloud. However, edge-based RL has not been widely studied so far. Instead, RL is often utilized as an aside system to Edge Computing, to help service offloading, task scheduling, or resource allocation \citep{yao2022edgecloud}. Therefore, our work can be regarded as the pioneering study on edge-based RL, with the slow-fast collaborative study mechanism between cloud and edge, similar to \citep{Madan2021metaRim} and \citep{chen2021mc2sf}. To differentiate from the traditional edge AI methods, we use the notation `mobile' instead of `edge' since most impressions of internet RS are on the users' mobile devices (smartphones or pads). 

\section{Preliminary and Problem Formulation}
\label{sec:preliminaries}

This section illustrates key concepts of our approach, including the system configuration, formulation, and detailed structure of mccHRL, as well as some necessary preliminaries.

\subsection{Listwise Recommending Scenario}

There is assumed to be a listwise recommendation task that exhibits $K$ items upon each user query, in which the session length $K$ is a pre-determined, fixed integer. User clicks response to exposed items within a session is then $\mathbf{c} = \{c_0, c_1, \cdots, c_K \}$ in which each $c_k$ is a binary variable, $k \in [0, K]$. The recommendation objective is generally to maximize the global session-wise click-through rate (CTR).

We first encoder each item into the embedding vector $e \in \mathbb{R}^L$. Then a session recurrent encoder (SRE) is employed to encode the session's historically exposed item sequence (We simply use the notation `historical sequence' in the following contexts for briefly) into the same embedding space, $l := \{ e_0, e_1, \cdots, e_K \} \in \mathbb{R}^L$. More details of SRE will be introduced in Section \ref{sec:sre}.

For personalization purpose, the user general profile, the user on-device feature, and the STA information are encoded into $u$, $m$ and $c^o$. respectively. The superscript $o$ denotes the outra-session context, to differentiate from the intra-session context ($c^i$) which is encoded inside the actor and will be mentioned in the later contexts.

\subsection{Reinforcement Learning}
\label{sec:rl}
Reinforcement Learning (RL) is an interactive learning technology that is generally built on Markov Decision Process (MDP). MDP can be represented with a four-tuple of $\mathcal{M}_t := (\mathcal{S}, \mathcal{A}, \mathcal{R}, T)$, where $\mathcal{S}$ is the set of state $s$, $\mathcal{A}$ is the set of action $a$, $\mathcal{R}$ is the set of reward $r$, and $T(s_{t+1} \vert s_t, a_t)$ is the transition function of $s_{t+1}$ after executing $a_t$ on $s_t$. The subscript $t$ of $\mathcal{M}$ works as the step indicator. 

RL optimizes a long-term objective which is defined as the discounted accumulated rewards $J = \sum_{t=0}^{\infty} \gamma^t r_t$, where $\gamma \in [0, 1)$ is the discount factor. The goal of reinforcement learning is to learn a policy $\pi (a | s)$ which maximizes $J$. In this work, we employ the classical model-free and off-policy algorithm called the Deep Deterministic Policy Gradient (DDPG) method \citep{lillicrap2019continuous} to solve this problem.


\begin{figure*}[t]
  \centering
  \includegraphics[width=1.0\columnwidth]{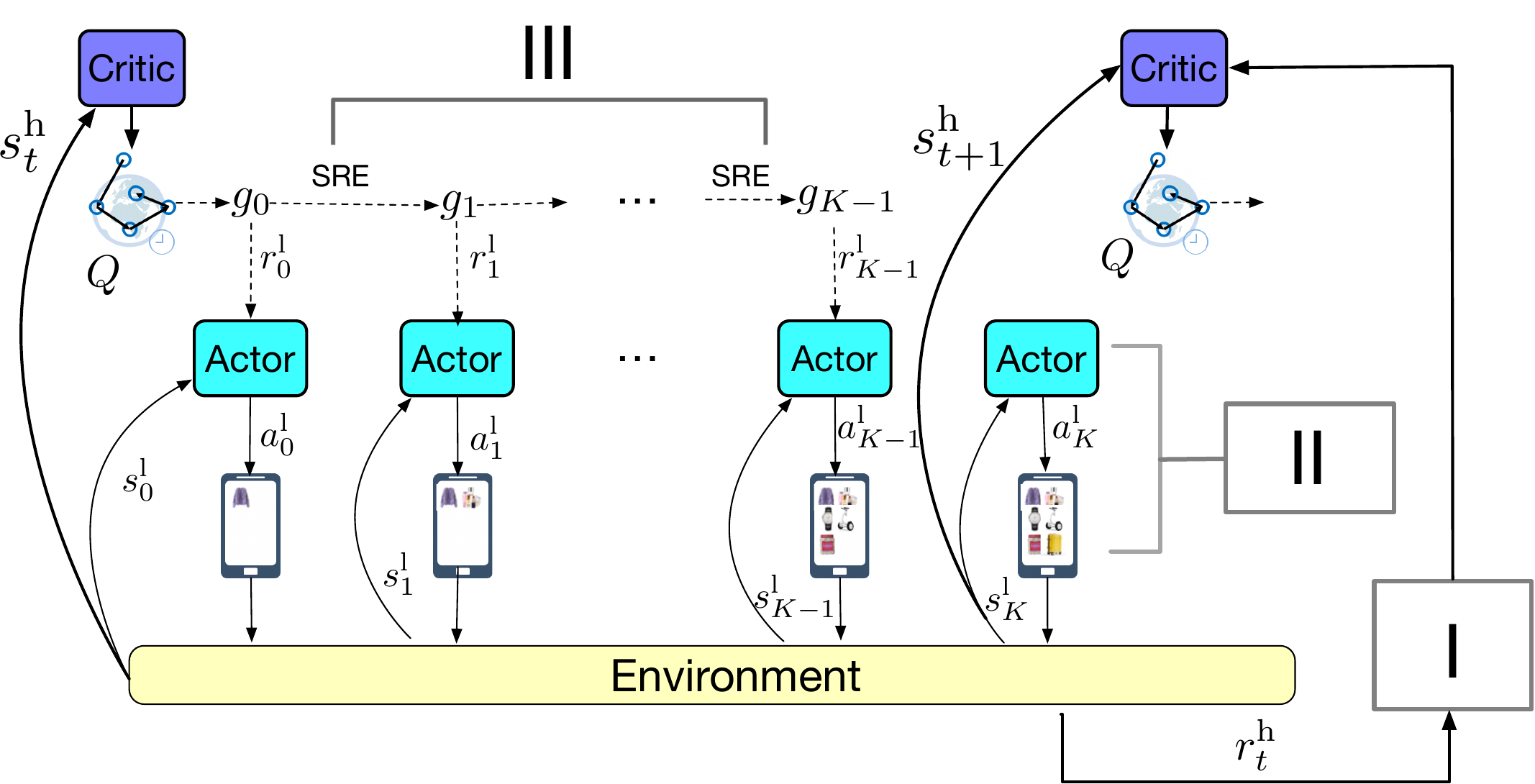} 
  \caption{The interaction procedure of mccHRL. I: Off-policy training of HRA, with external reward collected at the end of session. II: On-policy training of LRA, with intrinsic motivation from the goal. III: Goal initiates from the action of HRA, with a transit function provided by SRE implicitly along the session length.}
  \label{fig:hrl}
\end{figure*}

\subsection{Deep Deterministic Policy Gradient}
\label{sec:ddpg}

In this paper, we implement RL with a classical model-free and off-policy algorithm called the Deep Deterministic Policy Gradient (DDPG) method \citep{lillicrap2019continuous}. Here we briefly review its algorithmic details which is the basis of later derivation.

The DDPG is borrowed to get the recommended score of the item (\emph{i.e.}, action) and the long-term estimate return (\emph{i.e.}, Q-value).

DDPG has a typical actor-critic architecture in which the actor is the parametric policy network $\pi_{\theta}(a|s)$ while the critic is a state-action value function 
\begin{equation}
\label{eq:value_function}
    Q_w(s, a) = \sum_{i=t}^{\infty} \gamma^{i-t} E(r_i \vert s_t = s, a_t = a)
\end{equation}
in which $\theta$ and $w$ are trainable parameters. DDPG also keeps a target actor and a target cricic network with their trainable parameters $\theta^{'}$ and $w^{'}$ as instead.

With target networks fixed, $w$ can be updated by minimizing
\begin{equation}
\label{eq:ddpg_critic_bellman}
    y_t = r_t + \gamma Q_{w^{'}}(s_{t+1},\pi_{\theta^{'}}(s_{t+1}))
\end{equation}
then $w$ is updated by minimizing 
\begin{equation}
\label{eq:ddpg_critic_opt}
\small
L = \mathbb{E}_{s \backsim d^{\pi}}(r_t + \gamma Q_{w^{'}}(s_{t+1},\pi_{\theta^{'}}(s_{t+1})) - Q_{w}(s_{t},\pi_{\theta}(s_{t})))^2
\end{equation}
which is the famous Bellman Equation. $\theta$ is updated by the policy gradient
\begin{equation}
\label{eq:ddpg_actor_opt}
\triangledown_{\theta} J = \mathbb{E}_{s \backsim d^{\pi}} \triangledown_{a}Q_w(s,a) |_{a=\pi_{\theta}(s)} \triangledown_{\theta} \pi_{\theta}(s)  
\end{equation}
where $d^{\pi}(s)$ is the discounted distribution of state $s$ sampling by policy $\pi$. Target network parameters are softly updated  by 
\begin{align}
\label{eq:ddpg_target}
\theta^{\mu^{'}} {\leftarrow} \tau {\theta^{\mu^{'}}} + (1-\tau)\theta^{\mu} \notag \\
w^{'} \leftarrow \tau w^{'} + (1-\tau)w 
\end{align}
in a pre-defined time interval with $\tau \in (0, 1)$.

Then $\theta$ can be updated as 
\begin{equation}
\theta \leftarrow \theta + \eta \mathbb{E}_{s \backsim d^{\pi}}[\triangledown_{\theta} \pi_{\theta}(s)  \triangledown_{a}Q_w(s,a)|_{a=\pi(s)}]
\end{equation}
with $\eta$ as the learning rate. 


\subsection{Gated Recurrent Unit}
\label{subsec:gru}
Gated Recurrent Unit (GRU) is a special type of RNN that can sequentially capture the item-dependency impacts. It is found that GRU has better stability with fewer parameters than general GRN, therefore is usually used in recommendation studies \citep{Hidasi2016GRUrec, zhao2019LIRD, zhou2019edge}. The core logic of GRU is the following:
\begin{equation}
\label{eq:gru}
\begin{split}
    & \mathnormal{z}_{t} = \mathnormal{\sigma} (W_z \cdot [h_{t-1}, x_t]) \\
    & \mathnormal{r}_{t} = \mathnormal{\sigma} (W_r \cdot [h_{t-1}, x_t]) \\
   & \tilde{h}_t  = \mathnormal{tanh} (W_h \cdot [\mathnormal{r}_{t} h_{t-1}, x_t]) \\
   &  h_t =(1 - z_t) h_{t-1} + z_t (\tilde{h}_t)
\end{split}
\end{equation}
where $\mathnormal{z}_{t}, \mathnormal{r}_{t}$ are update and reset gates.



\subsection{The mccHRL Framework}

Recalling the discussion in Section \ref{sec:related_work}, the listwise ranking problem can be defined as two MDPs in varied time scales:
\begin{enumerate}
    \item The user state evolution at the real physical time $t$, with the notation of MDP as $\mathcal{M}_t$;
    \item Determination of the $k$th optimal item selection among remained $K-k$ candidates. The MDP is notated by $\mathcal{M}_k$, with $k$ as the virtual decision step. 
\end{enumerate}
In our work, we use HRL to address $\mathcal{M}_t$ and $\mathcal{M}_k$ simultaneously. We define our High-Level recommendation Agent (HRA) as the user state encoder, with the tuple $(s^{\text{h}}, a^{\text{h}}, r^{\text{h}})$ in $\mathcal{M}_t$; while the Low-Level recommendation Agent (LRA) is the item selector, working on $\mathcal{M}_k$ with tuple $(s^{\text{l}}, a^{\text{l}}, r^{\text{l}})$. Below are further detailed definitions:
\begin{itemize}

    \item \textbf{High-Level State $s^{\text{h}}$}: the user preference including the user profile $u$, the user-wise browsing session histories $\mathbf{l} := \{l_1, l_2, \cdots, l_{N^l} \}$, and the outra-session context $c^o$.
    \item \textbf{High-Level Action $a^{\text{h}}$}: embedding of the favorable recommended session $l \in \mathbb{R}^L$. 
    \item \textbf{High-Level Reward $r^{\text{h}}$}: the user CTR response w.r.t the current session, \textit{i.e.}, $\sum \mathbf{c}/K$.
    \item \textbf{Low-Level State $s^{\text{l}}$}: past item-selection decisions within a session inference. It includes the user on-device feature $m$, the intra-session context $c^i$, the current selected item sequence $\{e_1, e_2, \cdots, e_{k} \}$, and also the user profile $u$.
    \item \textbf{Low-Level Action $a^{\text{l}}$}: the studied user local, preference $\hat{e}_k \in \mathbb{R}^L$, as an indicator for item selection. 
    \item \textbf{Low-Level Reward $r^{\text{l}}$}: the intrinsic motivation received by LRA. Here we embrace the goal-conditioned framework of HRL \citep{Vezhnevets2017FuN, Nachum2018HIRO, Nasiriany2019LEAP}, \textit{i.e.}, the nearest action of HRA studies the current user preference and therefore can be used to calculate the goal of LRA action. For instance, the LRA reward $r^{\text{l}}$ can be simply formulated as the Ellucidian distance between the HRA action and LRA action.
    
\end{itemize}

Objectives of HRA and LRA are then formulated as
\begin{align}
\label{eq:v}
   J^{\text{h}}(a^{\text{h}}) &= \sum_{t=0}^{\infty} \gamma^{t} r^{\text{h}}_{t}, \quad J^{\text{l}}(a^{\text{l}}) = \sum_{k=0}^{K} \gamma^{k} r^{\text{l}}_{k}
\end{align}
in which the definition of the LRA reward connects the above two equations, \textit{i.e.}, $r^{\text{l}}(a^{\text{h}}, a^{\text{l}})$.

\section{Method}
\label{sec:method}


\begin{table}[t]
 \caption{Notations.}
 \label{tab:note}
 \centering
  \begin{tabular}{cl}
    \hline
    Notation & Description\\
    \hline
    $t$ & the real physical time \\
    $k$ & the position \& decision step within the recommended item list \\
    $K$ & the maximum length of recommended item list \\
    $\mathcal{S}$, $s$ & state space $\mathcal{S}$, $s \in \mathcal{S}$ \\
    $\mathcal{A}$, $a$ & action space $\mathcal{A}$, $a \in \mathcal{A}$ \\
    $\mathcal{R}$, $r$ & reward space $\mathcal{R}$, $r \in \mathcal{R}$ \\
    $\gamma$ & discount factor to balance immediate and future rewards \\
    $J$ & discounted cumulative return\\
    $\mathcal{T}$ & the transition $(s_t, a_t, r_t, s_{t+1})$\\
    $\mathcal{M}$ & MDP formed by $(\mathcal{S}, \mathcal{A}, \mathcal{R}, \gamma)$ \\
    $\pi_{\theta}(a|s)$ & policy function, parameterized by $\theta$\\
    $Q_{w}(s, a)$ & state-action value function, parameterized by $w$\\
    $c \in [0, 1]$ & binary response of user click \\
    $L$ & embedding dimension\\
    $u \in \mathbb{R}^L$ & embedding of user feature \\
    $c^o \in \mathbb{R}^L$ & embedding of outra-session context \\
    $c^i \in \mathbb{R}^L$ & embedding of intra-session context \\
    $m \in \mathbb{R}^L$ & embedding of on-device user profile \\
    $e \in \mathbb{R}^L$ & embedding of item feature \\
    $l \in \mathbb{R}^L$ & embedding of session (sequence of items within the session) \\
    $SRE$ & encodes any length of item sequence to $\mathcal{R}^L$ \\
    $g$ & the studied user long-term preference by HRA, the goal of LRA \\
    $\hat{e}$ & the studied user short-term preference by LRA \\
    $\hat{k}$ & index of the optimal item to exhibit at the next position \\
    $\mathscr{R}$ & the replay memory buffer \\
    $N$ & the batch size of off-policy sampling \\
    $\eta$ & the learning rate \\
    $\tau$ & the learning weight of target critic and actor \\
    $\alpha$ & the weighting parameter of CQL correction term \\
  \hline
\end{tabular}
\end{table}


This section introduces the training mechanisms of mccHRL, with Table \ref{tab:note} summarizing important symbols frequently used in the paper. We inherite the famous actor-critic RL architecture. However, here we let the HRA to be the critic and the LRA to be the actor solely, to reduce the computational complexity.


\begin{figure*}[t]
  \centering
  \includegraphics[width=1.0\linewidth]{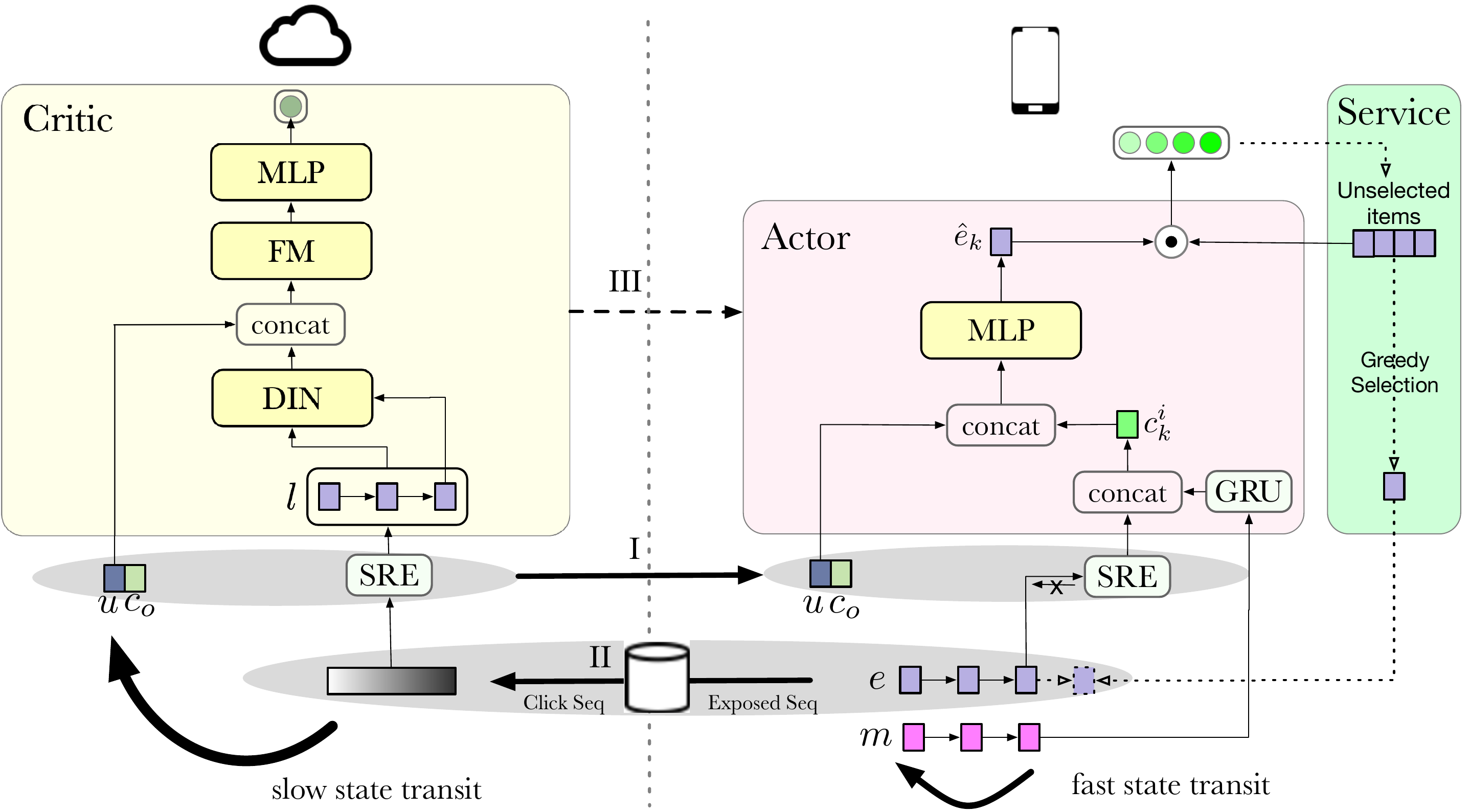}
  \caption{The Network Structure as well as Data Flows. 
  Arrow I: embedding vectors downloaded from Critic to Actor at the beginning of the session;
  Arrow II: user click behaviors uploaded with latency; 
  Arrow III: HRA action called for intrinsic motivation of LRA.} 
  \label{fig:actor_critic}
\end{figure*}

\subsection{Session Recurrent Encoder}
\label{sec:sre}

The session recurrent encoder (SRE) is an incremental function that aims to provide a session embedding with arbitrary session length and can be applied both in the learning and planning stages. Here we utilize the famous GRU structure, with $e_k$ as the input:
\begin{equation}
    l_{k+1} = \text{SRE} (l_k, e_k), \quad k = 0, \cdots, K-1
\end{equation}
the encoded session embedding $l_k$ is the latent state of GRU which has the same dimension of $e_k$.

\subsection{High-Level Agent}

The HRA performs as the critic in our mccHRL. It is modeled with a parametric value function $Q_{w}(s^{\text{h}}, a^{\text{h}})$. The left part of Figure \ref{fig:actor_critic} shows the critic network structure. Historical click sequences $\{ e \}$ are extracted from the cloud database, each encoded by SRE and then processed by a DIN structure \citep{Zhou2018DIN} with the last $l$ as the target tensor. The DIN output concatenates with $u$ and $c^o$ and goes into first an FM layer first then an MLP. The last activation function of MLP is tanh which generates the Q-value. At the beginning of a session service, HRA also transmits $u$, $c^o$, and SRE to LRA, as indicated by Figure \ref{fig:actor_critic}. The output of $Q_w$ then provides an implicit function of LRA goals $g$, as indicated by Arrow III.




\subsection{Low-Level Agent}

Figure \ref{fig:actor_critic} shows the actor-network structure on the right, which implements our LRA policy function $\pi_{\theta}(a^{\text{l}})|s^{\text{l}}))$. Since the size of the edge model is significantly limited by the on-device hardware, we design a relatively simple network structure. The local intra-session exposed item sequence $\{e_0, e_1, \cdots, e_k \}$ is again encoded by SRE, while the on-device user profile sequence $\{m_0, m_1, \cdots, m_k \}$ is encoded by another GRU unit. Concatenation of the last latent states of SRE and GRU forms the intra-session context $c^i$. We further concatenate $u$, $c^o$ and $c^i$ together and feed into a MLP which generates $\hat{e}_k \in R^L$, as $a^{\text{l}}_k$. During the training stage, gradient back-propagation of SRE in the actor is blocked to reduce the computational complexity.



During the service stage, $\hat{e}_k$ is dot multiplied with the embeddings of unselected items and generates scalar scores, then greedily selects the optimal item:
\begin{equation}
\label{eq:select}
    \hat{k} = \arg \max \{ \hat{e}_k e_{k^{\prime}}^T \}, \quad k^{\prime} = k+1, \cdots, K^{\prime}
\end{equation}
in which $K^{\prime}$ is the remained item candidate set size (and can be assumed to be larger than $K$). The $\hat{k}$th item is appended to the end of recommendation, then the step moves forward from $k$ to $k+1$ until all $K$ items are determined.

\subsection{Mechanism}


Figure \ref{fig:hrl} reviews the aforementioned interactive pattern of HRA and LRA. HRA operates at the real world time $t$, observes the current user state $s^{\text{h}}_t$, and provides the user favorable preference by the High-Level action $a^{\text{h}}_t$ once a session service is requested. The HRA training is executed on the cloud side based on all users' logs, and policy updates once the session ends with the real, non-delayed session-wise user response $r^{\text{h}}_t$. Within each session's servicing, LRA aims to determine the ranked list of $K$ items, therefore operating at the virtual time steps $k= 0, 1, \cdots, K$  on the device side. 

\subsubsection{Training with Generic Environment}

Algorithm \ref{alg:mcchrl} shows the training mechanism of our mccHRL. Considering the listwise mechanism, We assign the Low-Level actor an on-policy strategy and the High-Level critic an off-policy strategy. Similar to DDPG, we also have a target critic and a target actor.



\subsubsection{Training with Offline Data}

Algorithm \ref{alg:mcchrl} is an interactive learning method with both cloud and edge environments. Unfortunately, it is often not practical for direct online deployment and learning on large-scale industrial systems, especially for online learning on users' mobile devices. Therefore, it is of great importance to develop an algorithm that can learn from the offline data, or at least use the offline training as a warmup start. In this paper, we assume offline data can be retrieved and reformulated in the temporal order, with each line in the form of $[s^{\text{h}}_t, a^{\text{h}}_t, r^{\text{h}}_t, \{ e_k \}, \{ m_k \}, s^{\text{h}}_{t+1}, a^{\text{h}}_{t+1}]$.

RL study on such data calls for a special research direction called offline RL, or batch-constrained RL, which learns an optimal policy from a fixed dataset by implementing the out-of-distribution corrections for off-policy RL \citep{levine2020offline}. Since our HRA has an off-policy strategy, we could employ similar ideas to train the HRA critic offline. Here we refer to the conservative Q-learning (CQL) \citep{Kumar2020CQL}, to have a regularizer in the TD loss of the Q-function:
\begin{align}
    L_{\text{off}}(s, a, \pi_{\theta}) = &\mathrm{E}_{a \sim \pi_{\theta}} [Q_w(s, a) \frac{\exp{Q_w(s, a)}}{Z}] \notag \\
    &- \mathrm{E}_{a \sim \pi_{\beta}} Q_w(s, a) \label{eq:Loff}
\end{align}
in which $Z$ is the normalizing factor of $Q$ and $\pi_{\beta}$ is the sampling policy in data. This formulation indicates that we also minimize the soft-maximum of Q at the current solved policy ($\pi_{\theta}(a|s)$) w.r.t the sampling policy $a^{\text{h}}$, therefore the learning becomes 'conservative'. According to Algorithm 1 in \citep{Kumar2020CQL}, such a trick can be also applied in the actor-critic framework. Here we present our offline version of Algorithm \ref{alg:mcchrl} which is specified in Algorithm \ref{alg:mcchrl_offline}.


\begin{algorithm}[ht!]
\caption{The mccHRL Algorithm.}
\label{alg:mcchrl}
\begin{algorithmic}[1]
\STATE \textbf{Initialize} the clock $t \leftarrow 0$
\STATE {\textbf{Initialize} parameters $w \leftarrow 0$,$\theta \leftarrow 0$, and $w^{\prime} \leftarrow w$, $\theta^{\prime} \leftarrow \theta$}
\STATE \textbf{Initialize} all users' histories $\mathbf{l} = \{ \}$
\STATE \textbf{Initialize} the on-cloud replay memory buffer $\mathscr{R} = \{ \}$
\STATE \textbf{foreach} \textit{session} \textbf{do} \\

\qquad // Data Transmission Stage 1, High-Level to Low-Level:

\STATE \qquad Observe $u_t$, $c^o_t$ and $s^{\text{h}}_t = [u_t, c^o_t, \mathbf{l}]$


\STATE \qquad \textbf{Initialize} $k \leftarrow 0$, locate the mobile device
\STATE \qquad Synchronize $u_t$, $c^o_t$, and candidates $e$ from cloud 
\STATE \qquad Synchronize $\theta \leftarrow \theta^{\prime}$ from cloud

\qquad // The Low-Level Training Stage:
\STATE \qquad \textbf{while} $k < K$ \textbf{do}
\STATE \qquad \qquad \textbf{Initialize} the ranked item list $\mathbf{e} = \{ \}$
\STATE \qquad \qquad Collect the instant $m_k$ and record $s^{\text{l}}_k = [u, c^o, m_k, \mathbf{e}]$
\STATE \qquad \qquad Get $a^{\text{l}}_k$ based on the current policy $\pi_{\theta}(a^{\text{l}}_k \vert s^{\text{l}}_k)$
\STATE \qquad \qquad Select item $\hat{k}$ from Eq. (\ref{eq:select}) and append $\mathbf{e}$ with $e_{\hat{k}}$
\STATE \qquad \qquad Update $\theta$ based on DDPG 

\STATE \qquad \textbf{end while}

\qquad // Data Transmission Stage 2, Low-Level to High-Level:

\STATE \qquad \textbf{With some latency}:
\STATE \qquad \qquad Update the target actor $\theta^{\prime} \leftarrow \tau \theta + (1 - \tau) \theta^{\prime}$ 
\STATE \qquad \qquad Collect $\mathbf{c}$, calculate $r^{\text{h}}_t$, and append $\mathbf{l}$ with $\mathbf{e}$. 
\STATE \qquad \qquad Observe the next HRA state $s^{\text{h}}_{t+1}$
\STATE \qquad \qquad Store transition $\mathcal{T}_t = (s^{\text{h}}_t, a^{\text{h}}_t, r^{\text{h}}_t, s^{\text{h}}_{t+1})$ in $\mathscr{R}$

\qquad // The High-Level Training Stage:
\STATE \qquad \textbf{if} $\text{size}(\mathscr{R}) > N$ \textbf{then}
\STATE \qquad \qquad {Sample $N$ transitions $\mathcal{T}_i$ from $\mathscr{R}$}
\STATE \qquad \qquad \textbf{for} $i = 1, \cdots, N$ \textbf{do}
\STATE \qquad \qquad \qquad Update $w$ based on DDPG 
\STATE \qquad \qquad Update the target critic $w^{\prime} \leftarrow \tau w + (1 - \tau) w^{\prime}$
\STATE \qquad \textbf{end if}

\STATE \textbf{end for}
\end{algorithmic}
\end{algorithm}

\begin{algorithm}[!htb]
\caption{The offline mccHRL Algorithm.}
\label{alg:mcchrl_offline}
\begin{algorithmic}[1]
\STATE \textbf{Initialize} parameters $w \leftarrow 0$,$\theta \leftarrow 0$
\STATE {\textbf{Initialize} the replay memory buffer $\mathscr{R} = \{ \}$} 
\STATE \textbf{foreach} \textit{line of data} \textbf{do} \\

\STATE \qquad Retrieve $s^{\text{h}}_t, a^{\text{h}}_t, r^{\text{h}}_t, \{ e_k \}, \{ m_k \}, s^{\text{h}}_{t+1}, a^{\text{h}}_{t+1}$ from data



\STATE \qquad Store transition $\mathcal{T}_t = (s^{\text{h}}_t, a^{\text{h}}_t, r^{\text{h}}_t, s^{\text{h}}_{t+1}, a^{\text{h}}_{t+1})$ in $\mathscr{R}$

\qquad // The Cloud Training Stage:
\STATE \qquad \textbf{if} $\text{size}(\mathcal{R}) > N$ \textbf{then}
\STATE \qquad \qquad {Sample $N$ transitions $\mathcal{T}_i$ from $\mathscr{R}$}
\STATE \qquad \qquad \textbf{for} $i = 1, \cdots, N$ \textbf{do}
\STATE \qquad \qquad \qquad {Set $y_i = r^{\text{h}}_i + \gamma Q_{w}(s^{\text{h}}_{i+1}, a^{\text{h}}_{i+1})$}
\STATE \qquad \qquad \qquad Calculate $L$ based on in Eq. (\ref{eq:ddpg_critic_opt})
\STATE \qquad \qquad \qquad Calculate $L_{\text{off}}$ based on in Eq. (\ref{eq:Loff})
\STATE \qquad \qquad \qquad Update $w$ by $\min L + L_{\text{off}}$
\STATE \qquad \textbf{end if}

\qquad // The Edge Training Stage:
\STATE \qquad \textbf{for} $k = 0, \cdots, K$ \textbf{do}
\STATE \qquad \qquad \textbf{Initialize} the ranked item list $\mathbf{e} = \{ \}$
\STATE \qquad \qquad Observe $m_k$ and $s^{\text{l}}_k = [u_t, c^o_t, m_k, \mathbf{e}]$
\STATE \qquad \qquad Get $a^{\text{l}}_k$ based on the current policy $\pi_{\theta}(a^{\text{l}}_k \vert s^{\text{l}}_k)$
\STATE \qquad \qquad Observe the sampled item $e_k$ and append it to $\mathbf{e}$
\STATE \qquad \qquad Update $\theta$ by $\triangledown_{a}Q_w(s^{\text{h}}_t, a^{\text{l}}_k) \triangledown_{\theta} \pi_{\theta}(s^{\text{l}}_k)$
\STATE \qquad \textbf{end for}

\STATE \textbf{end for}
\end{algorithmic}
\end{algorithm}

\section{Experiments}
\label{sec:experiment}

As an on-policy RL-based methodology, our mccHRL should be evaluated by online experiments and business performance indicators such as CTR. However, the industrial recommendation system is complex and connected with enormous impressions and incomes, which makes thrugh online experiment difficult. As a remedy, we design two offline evaluation strategies. We demonstrate the effectiveness of mccHRL by both offline and live experiments, as well as ablation and sensitivity studies.

\subsection{Experimental Environments}

In order to evaluate our RL-based method, ideally we need a simulator with both cloud and edge environments, which are interactively learnable by RL agents. Unfortunately, to the best of our knowledge, there is no such test-purpose cloud-edge request simulator. Instead, we provide two remedy solutions for offline experiments, before the online deployment:

\begin{enumerate}
    \item An RS simulator that is generated by a small-scale public dataset. We arbitrarily define some spatial and temporal related features as `edge features' which is not observable by the cloud. Other features are manually implemented with some transmission latency between the edge and the cloud. The average reward during the test stage can be used as a metric for such a simulation-based system.
    \item Experiment based on a large-scale industrial dataset. The offline version of Algorithm \ref{alg:mcchrl} can be employed on such an offline RL (or dataset-based RL) problem, with some offline accuracy metrics to be evaluated.
\end{enumerate}


We employ both two solutions to evaluate our methodology. Implementation details are introduced in the subsequent subsections. Table \ref{tab:stat} exhibits the statistics of two experimental environments.

\subsubsection{Simulator-based Experiment.}

Movielens \footnote{https://grouplens.org/datasets/movielens/} is a user-movie rating dataset including the behavior data, the user feature, and item feature data. Samples are labeled with users' ratings on movies from 1 to 5. 

Based on the MovieLens-100k dataset, We build a simulator based on the methodology introduced in \citep{zhao2019LIRD}, with the averaged ratings behaving as the RL rewards. The average rating is 3.53 over all samples. Among the user features, the spatial features including `occupation' and `zip code' are set as edge features, which can not be accessed by the cloud. We further assume the user cloud sequence has a fixed latency with edge sequence, \textit{i.e.}, the cloud sequence is delayed by 6 items the edge sequence. By these configurations, we build an approximated mobile-cloud simulator, with an explicit transmission latency.


\subsubsection{Dataset-based Experiment. }
Because of the industrial limitations, interactive training of on-device RL is temporarily impractical. Instead, we conduct an industrial-scale experiment based on the offline dataset with realistic on-device features. Here we employ a content-based feed recommendation dataset \footnote{https://tianchi.aliyun.com/dataset/dataDetail?dataId=109858}, sampling from the Taobao front-page content-base recommendation and has been declared open-sourced in Tianchi. On-device features are aligned with the recommended item and corresponding click labels, and the data is reorganized into a session-wise form. The recommendation objective is to maximize the click-through rate (CTR). We denote user app-related behaviors sampled on the device as $m$, including app IDs, stay-times, and LBS information in the granularity of districts. The spatiotemporal information in $c^o$ including POI, province, city, day, hour, and workday/weekend labels. We further assume the user cloud sequence has a fixed latency with edge sequence, \textit{i.e.}, the cloud sequence is delayed by 10 items the edge sequence. 




\begin{table*}[!htb]
\caption{Statistics of datasets.}
\label{tab:stat}
\centering
\begin{tabular}{c|c|c|c|c|c}
\toprule
    Dataset & set & \#impression & \#user & \#item & CTR \\ 
\midrule
\multirow{6}{*}{Movielens} & All & 100000 & & & \\
& rate=1 & 6110 & & &  \\
& rate=2 & 11370 & 943 & 1602 & - \\
& rate=3 & 27145 & & & \\
& rate=4 & 34174 & & & \\
& rate=5 & 21201 & & & \\
\midrule
\multirow{3}{*}{Alibaba} & All & 265003 & & & 4.92\% \\
& day=3$\sim$9 & 226508 & 34708 & 8 & 4.85\% \\
& day=10 & 38495 & &  & 5.37\% \\
\bottomrule
\end{tabular}
\end{table*}

\subsection{Evaluation Metrics}


We define different evaluation metrics according to the characteristics of two experimental environments:
\begin{itemize}
    \item Simulator-based Experiment.: the expected reward indicates the policy performance in an interactive environment. The rating employed for evaluation is retrieved from the most similar occurrence (with a cos similarity by user and item embeddings) in the dataset. Such test stage is repeated for 50 rounds and the final averaged metric is reported. We evaluate it by the average rating during the test stage of the experiment, which is denoted by \textit{S-$<$rating$>$}.
    
    \item Dataset-based Experiment.: the general solution is to convert the online CTR maximization into an offline binary classification problem, \textit{i.e.} predict if the user clicks or not. We use Area Under the ROC Curve (AUC) to evaluate this classification performance, in which we employ the actor's servicing score as the prediction and the ground-truth click record as the label. We name it as \textit{D-AUC}.
\end{itemize}

\subsection{Implementation Details}
\label{sec:exp_detail}

Table \ref{tab:hyperparam} shows some hyper-parameters of two experiments. Further implementation details are listed below.

\begin{table*}[!htb]
  \caption{Parameter Settings.}
  \label{tab:hyperparam}
  \begin{center}
  \begin{tabular}{ccc}
    \hline
    Hyper-parameter & Simulator-based Experiment & Dataset-based Experiment \\
    \hline
    $K$ & 4 & 6 \\
    $\gamma$ & 0.99 & 0.9 \\
    $L$ & 128 & 16 \\
    $N$ & 64 & 512 \\
    $\eta$ & 0.001 (HRA), 0.0001 (LRA) & 0.0001 \\
    $\tau$ & 0.001 & 0.001 \\
    $N^l$ & 12 & 50 \\
    $\alpha$ & - & 0.1 \\
  \hline
\end{tabular}
\end{center}
\end{table*}

\subsubsection{Simulator-based Experiment.}


MLP in critic has hidden units of [32, 16, 1] with the last activation as tanh; while MLP in actor has hidden units of [128, K]. The latent state dimension is 16 for both GRU and SRE components.  We use an Adam optimizer to interactively train the critic and actor, with a learning rate of 0.0001. Performance is evaluated by iterating all users among the training set, while for each ground truth item, the policy selects top-$K$ items based on scores. Because the size of MovieLens-100k is relatively small, we omit the DIN structure here to avoid overfitting



\subsubsection{Dataset-based Experiment.}

The DIN layer of critic uses a multi-head target attention structure, with the last session embedding vector as the target. The attention mode is cosine. The deepFM layer of critic is composed of linear terms and 2nd-order cross terms. MLP in critic has hidden units of [128, 64, 32, 1] with the last activation as tanh; while MLP in actor has hidden units of [64, 32, K] with all activation as relu.  The latent state dimension is 16 for both GRU and SRE components. We use the Adam optimizer with a learning rate of 0.0001, 1st-regularization weight of 0.0001 and 2nd-regularization weight of 0.00001. Data on the last day (day = 10) is classified into the test set while the other consists of the training set. To help the training converge, we first conduct 3 epochs of supervised learning, with the binary cross-entropy loss of the click label. RL then conducts 2 more epochs. The training ends with convergence, costing about 17000 steps. Training is ruuning on an industrial-scale, parallel training platform, with each experiment allocated with 8 GPUs and 20000M RAM.

\subsection{Baselines}

The following experimental baselines are employed to compare with our mccHRL: 

\begin{itemize}
    \item \textit{Random}: the policy picks the recommended item from the candidates set randomly. Only implemented in the simulator experiment.
    \item \textit{DIN}: the Deep Interest Network \citep{Zhou2018DIN}, as a standardized solution of industrial pointwise CTR model.
    \item \textit{DIN}(ideal): the DIN model (which is deployed on the cloud) can have full access to edge features with exact zero latency. Therefore, it indicates an unrealistic performance upper bound for a generic supervised model. Only implemented in the industrial experiment.
    \item \textit{GRU4rec}: a session-based method with a latent state calculated by GRU \citep{Hidasi2016GRUrec}, therefore considers the intra-session context.
    \item \textit{LIRD}: a DDPG-based RL listwise ranking framework \citep{zhao2019LIRD}. 
    \item \textit{MC$^2$-SF}: a slow-fast training framework based on moble-cloud collaboration \citep{chen2021mc2sf}, with mobile features incorporated and considered.
\end{itemize}

\begin{table}[t]
  \caption{Offline Experiment Results}
  \label{tab:results}
\begin{center}
  \begin{tabular}{ccc}
\hline
    Experiment & S-$<$rating$>$  & D-AUC \\ %
\hline
    random & 3.583 $\pm$ 0.034 & - \\ 
    DIN & 3.767 $\pm$ 0.045 & 0.781 $\pm$ 0.006 \\
    DIN(ideal) & - & 0.832 $\pm$ 0.008 \\
    GRU4rec & 3.624 $\pm$ 0.038 & 0.731 $\pm$ 0.007 \\
    LIRD & 3.786 $\pm$ 0.051 & 0.631 $\pm$ 0.007\\
    MC$^2$-SF & 3.733 $\pm$ 0.050 & 0.769 $\pm$ 0.010 \\
\hline
    mccHRL & \textbf{3.824} $\pm$ 0.049 & \textbf{0.853} $\pm$ 0.009 \\
    mccHRL(wo edge) & 3.746 $\pm$ 0.046 & 0.787 $\pm$ 0.009 \\
    mccHRL(wo actor) & 3.770 $\pm$ 0.045 & 0.707 $\pm$ 0.008 \\
    mccHRL(wo critic) & 3.587 $\pm$ 0.048 & 0.676 $\pm$ 0.008 \\
\hline
\end{tabular}
\end{center}
\end{table}

\subsection{Offline Results}

We compare the offline experimental performance of mccHRL and baselines in Table \ref{tab:results}. mccHRL outperforms other baselines in both experiments. Specifically, the improvement of mccHRL on the industrial experiment is more evident, since this experiment has substantial on-edge features, and user behaviors are more responsible for session-wise exhibition and industrial computational latency.

Comparison with baselines could highlight more information. For example, LIRD is the second-best method in the simulator experiment since its RL nature fits the interactive environment. However, it does not perform well in the industrial experiment, probably because that it does not have the offline training correction. GRU4rec studies the intra-session context by a GRU unit but is not enough for our industrial experiment in which the outra-session context is also important. MC$^2$-SF could provide reasonable performance with consideration of mobile features but lack the RL logic compared with our mccHRL. Finally, DIN(ideal) has a similar performance with mccHRL, indicating that a SOTA, supervised learned model could also perform well with immediate access to edge features, as well as no edge-cloud transmission latency. However, this assumption is obviously impractical in the real world.







\subsection{Ablation Study}
Compared with previously published works, our mccHRL has several novel components therefore their ablation studies are necessary. Table \ref{tab:results} also illustrates the results of the following attempts:
\begin{itemize}
    \item mccHRL(wo edge): the mccHRL approach with only cloud-based features.
    \item mccHRL(wo actor): the mccHRL but without the actor. We simply use $\arg \max Q$ to determine the favorable item selection.
    \item mccHRL(wo critic): the mccHRL but without the critic. An item selection policy is trained offline and is deployed on the device for service.
\end{itemize}
Not surprisingly, mccHRL still has the best performance, suggesting that the critic, the actor, and edge-based features are all crucial. Note that mccHRL(wo critic) has the worst performance since it suffers from the sparse reward problem.

  
\subsection{Sensitivity Study}
\label{sec:sens}

There is always a trade-off between performance and speed for the edge-based model, due to limited resources on the edge side. To have a clear picture, we choose an important factor, the user behavior sequence length kept by the edge model, to have sensitivity analysis. We inspect its impact from the following two aspects:
\begin{enumerate}
    \item The latency between cloud and edge sequence update. Since the detailed value of time latency is highly subject to the cloud and edge real-time conditions, here we pick an approximate number, the number of items such that cloud is behind the edge sequence. With this number larger, LRA could maintain more information but with a more computational cost.
    \item The length of behavior sequence studied by $s^{\text{l}}$ in LRA. If this number is larger, LRA could provide more historical information to policy but the model size is expected to increase.
\end{enumerate}

\begin{figure}[htb]
  \centering
  \includegraphics[width=1.05\columnwidth]{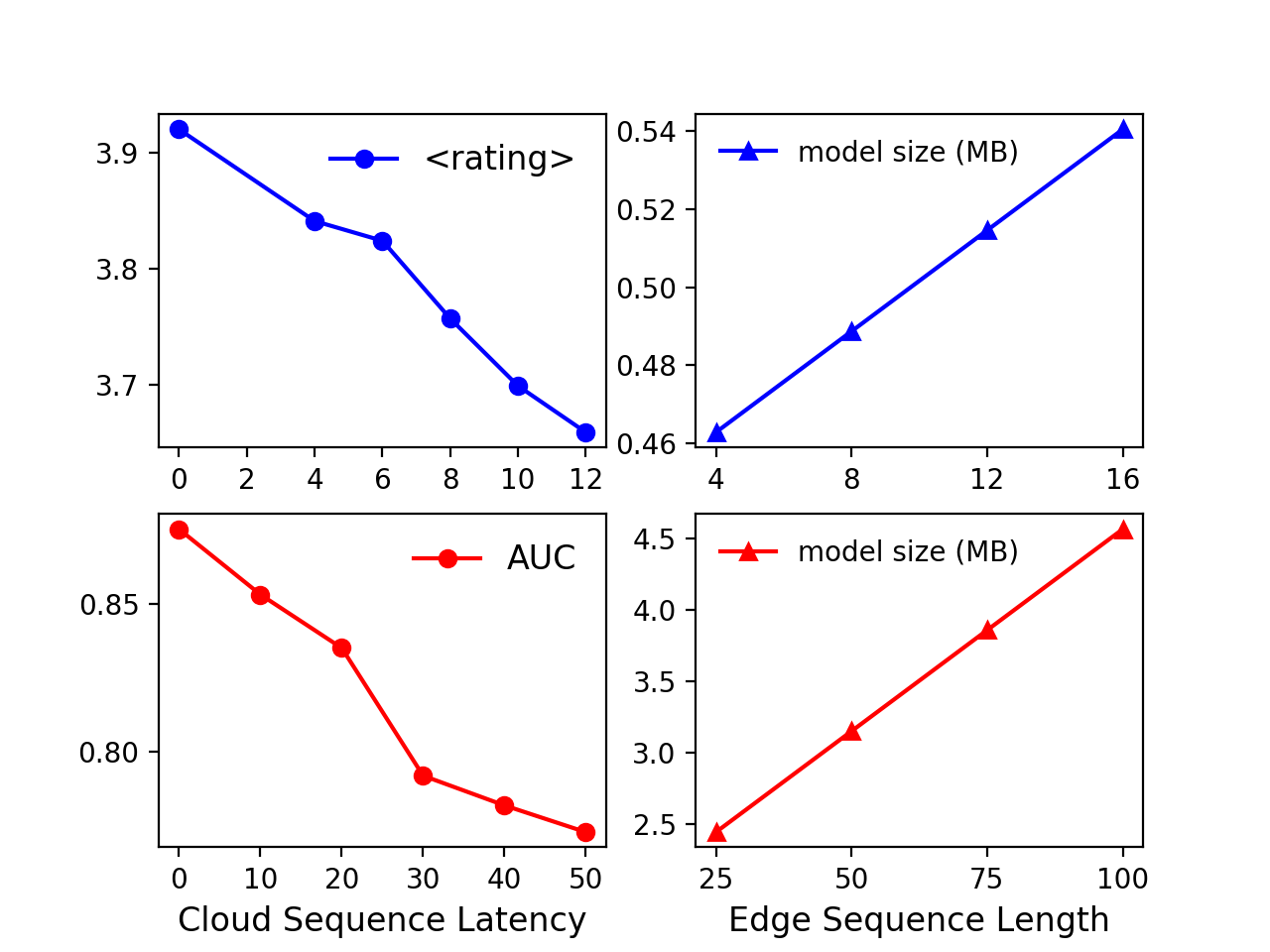} 
  \caption{Sensitivity plots of the Simulator experiment (blue) and the Industry experiment (red). Left: Plot of performance metrics with different cloud sequence latency. Right: the edge model size as a function of sequence length.}
  \label{fig:sens}
\end{figure}

Figure \ref{fig:sens} shows the sensitivity results of the above two factors. Not surprisingly, the model performance deteriorates as latency becomes larger according to Figure \ref{fig:sens} (Left), indicating that the edge model memory is a positive factor. However, improving the model memory will challenge the deployment ability on the device, as indicated by Figure \ref{fig:sens} (Right). Based on our practical experience, the edge model should generally be no bigger than 3.5MB to avoid unreasonable mobile overhead, which indicates our current choice of $N^l = 50$ is almost optimal.

\subsection{Live Experiment}
We deploy our algorithm on a world-leading mobile application and execute an A/B live test. The experiment lasts a week. We randomly select a pool of users from different geological areas to conduct our mobile-based deployment. Compared with the live baseline, we observe the global CTR increased by $1.2\%$ and the total views increased by $1.5\%$. 



\section{Discussion and Limitations}

This work is motivated by decoupling the two time scales of the listwise recommendation: the physical time step (for user preference update) and the virtual step (for the ranked item selection). Our methodology adopts the fast-slow system by learning the above two mechanisms simultaneously by the HRL framework. We also take advantage of the edge computing technique by implement the low-level agent on the mobile device side, in which the localized and STA user information can be consider and feedback before uploading to the cloud. Therefore, our method provides the theoretical optimal solution for such recommending scenarios.

However, one potential drawback is the extra computational burden brought by the mccHRL framework. Here we perform a simple time complexity analysis for a brief discussion. Given the physical time steps as $T$ and the virtual time steps as $K$ (the session length), inference of a generic pointwise recommendation model requires $O(K * T)$, in which the model is called with the user and item pair independently. In our mccHRL, the time complexity would be $O(K^2 * T)$ since we need to calculate the session contextual information recurrently. When $K$ is too large (\textit{e.g. } larger than 100), mccHRL might be inefficient and impractical. Fortunately, in the industrial implementation, the user request is usually divided into sequential page requests. This solution reduces the actual $K$ therefore could alleviate the efficiency issue. The mobile-device transmission could also be a problem which is already discussed in Section \ref{sec:sens}.

Another important potential improvement is to consider the user's instant response within a session and make immediate adjustments for ranking results. Such considerations will add some user-interactive rewards for Low-Level RL, and our method then becomes a unified framework of cloud-based ranking and mobile-based reranking, which is a promising future direction.

\section{Conclusion}
\label{sec:conclusion}

In this paper, we build a unified framework to deal with the user preference mode and the listwise item ranking simultaneously, by a novel hierarchical RL methodology. We reinforce the state Markovness by edge features consideration and design a mobile-cloud collaboration mechanism. The goal-conditional mechanism is utilized to synchronize the user preference solved by high-level RL to the low-level RL. The outra- and intra-session contexts are decoupled and studied in two time scales.  We implement both a simulator and an offline industrial dataset to verify our methodology, before the formal live experiment. Our study shed some lights on the applications of HRL on listwise recommendation, and future studies can be developed to further implement RL interactively in large industrial environments.

\bibliographystyle{plainnat} 
\bibliography{main}


\end{document}